\documentclass{PoS}

\title{Dynamical Origin for the 125 GeV Higgs; a Hybrid setup}

\ShortTitle{Dynamical Origin for the 125 GeV Higgs}

\author{\speaker{Shaouly Bar-Shalom}\\
        Physics Department, Technion-Institute of Technology, Haifa 32000, Israel \\
        \email{shaouly@physics.technion.ac.il}}

\abstract{We describe a hybrid framework for electroweak symmetry breaking (EWSB), in which the Higgs mechanism is combined with a Nambu-Jona-Lasinio mechanism. The model introduces an unconstrained  scalar (i.e., acts as "fundamental" but not the SM field) and a strongly coupled doublet of heavy quarks with a mass around 500 GeV, which forms a condensate at a compositeness scale $\Lambda \sim {\cal O}(1)$ TeV. This setup is matched at that scale to a tightly constrained hybrid two Higgs doublet model, where both the composite and unconstrained scalars participate in EWSB. This allows us to get a good candidate for the recently observed 125 GeV scalar which has properties very similar to the Standard Model Higgs. The heavier (mostly composite) CP-even scalar has a mass around 500 GeV, while the pseudoscalar and the charged Higgs particles have masses in the range 200 -300 GeV.}

\FullConference{The European Physical Society Conference on High Energy Physics -EPS-HEP2013\\
		18-24 July 2013\\
		Stockholm, Sweden}

\begin{document}

\section{Introduction}
With the recent LHC discovery of the 125 GeV scalar particle
\cite{LHCHiggs},
we are one step closer to understanding the mechanism of EWSB.
Is it a Standard Model (SM) Higgs? or may be it is the SUSY light Higgs?
so far it certainly seems to resemble the SM Higgs,
but in the grand picture it makes no sense; where is the new physics
to account for the hierarchy problem, dark matter, flavor etc...?
there is still no hint for that and/or for SUSY yet.

Another attractive alternative for new TeV-scale physics
is strong dynamics. Albeit, the strong dynamics setup is difficult
to realize with a light (125 GeV) composite Higgs, unless this light
scalar state is a pseudo-Goldstone boson of a global symmetry
breaking at the strong interactions scale. Indeed, in this talk we
will toy with this idea, proposing a specific hybrid
framework for Dynamical EWSB (DEWSB), where an unconstrained
scalar field (which behaves as a ``fundamental" field) is added at the compositeness scale
where additional super-critical attractive 4-Fermi operators of heavy fermions form
a composite scalar sector \cite{ourhybrid}.
The ``fundamental" scalar is unconstrained
at the compositeness scale and may
result from the underlying strong dynamics,
e.g., it can be the pseudo-Goldstone boson mentioned above.
This strongly coupled composite-plus-fundamental sectors
are then matched at the compositeness scale
to a hybrid  2HDM with a 4th generation of heavy fermions 
(named here after h4G2HDM),
where the fundamental-like field ($\Phi_\ell$) couples
to the SM's lighter fermions and
the auxiliary (composite) field ($\Phi_h$) couples to the heavy 4th generation
fermions.\footnote{We do not consider here explicitly
the choice by which the 4th generation
leptons couple to the Higgs sector; they can either couple
to the fundamental Higgs or to the auxiliary field.
In either case, we assume that
their couplings are sub-critical and, therefore,
do not play any role in DEWSB (see also discussion in \cite{ourhybrid}).}

Let us recall an old idea:
that a heavy fermion, $\psi$, may be the agent for DEWSB
\cite{DEWSB-old,luty,hashimoto1,DEWSB-HLP,burdman1,hunghybrid}.
In this case, the Higgs is viewed as a fermion-antifermion bound state
$< \bar\psi \psi > \neq 0$, and
there is no need to introduce an elementary Higgs field.
One of the early attempts in this direction investigated
the possibility of using the top-quark
as the agent for DEWSB via top-condensation
\cite{Bardeen}, in a generalization
of the Nambu-Jona-Lassinio (NJL) model \cite{Nambu}.
However, the resulting dynamical top mass turns out to be
appreciably heavier than $m_t \approx 175$ GeV,
thus making it difficult for top condensation to provide a viable picture.
Moreover, top-condensate
models require the cutoff/threshold
for the new strong interactions to be of ${\cal O}(10^{17})$ GeV,
i.e., many orders of
magnitudes larger than $m_t$, thus
resulting in a severely fine-tuned picture of DEWSB.
Nonetheless, several interesting generalizations of the top-condensate
model which potentially avoid these obstacles, have been suggested. For example, the condensation of new heavier quarks
and/or leptons may drive EWSB \cite{DEWSB-old,luty,hashimoto1,DEWSB-HLP,burdman1,hunghybrid}.

\section{DEWSB with the NJL mechanism}

We adopt here a 
simple and ''modest" logic for parameterizing our ignorance with regard
to the details of the would be TeV-scale strong dynamics. In particular,
without any explicit model building, we follow two guiding principles
which underly the NJL model \cite{Nambu}:
\begin{enumerate}
\item Assume the existence of a strongly interacting fermionic
sector above the compositeness scale $\Lambda$.
\item Trade our ignorance for effective couplings +
appropriate boundary conditions at the scale $\Lambda$.
\end{enumerate}

In particular, at $\Lambda$, physics is described by an effective (attractive) 4-Fermi interaction of the strongly coupled fermions:
\begin{eqnarray}
{\cal L}_{NJL} = G_\psi \left(\bar\psi_L \psi_R \right)
\left(\bar\psi_R \psi_L \right)  ~.
\end{eqnarray}

One then solves the Gap equation (i.e., keeping only fermions loops - bubble diagrams), and if $G_\psi$ is greater than some critical value
$G_\psi > G_c$, then (see e.g., \cite{Bardeen}):
\begin{itemize}
\item EW symmetry can be broken
\item The field $\psi$ acquires a dynamical mass
\item The low-energy theory contains a scalar bound state: $S \sim <\bar\psi \psi >$
\end{itemize}

To get a realistic framework, one can introduce an auxiliary field $H$, which reproduces the 4-Fermi interaction when it is integrated out:
\begin{eqnarray}
{\cal L} = g_0^\psi \left(\bar\psi_L \psi_R H + h.c. \right) -
m_0^2 H^\dagger H \stackrel{integrate~H}{\longrightarrow}
{\cal L}_{NJL} = \frac{g_0^\psi}{m_0^2} \left(\bar\psi_L \psi_R \right)
\left(\bar\psi_R \psi_L \right)~.
\end{eqnarray}

Then, below the cutoff $\Lambda$, $H$ develops kinetic terms and
quartic interactions (from the fermion loops) in the effective action,
becoming a dynamical field, and the theory containing $H$ is
exactly equivalent to the theory written
in terms of the fermions with
$G_\psi = \frac{g_0^\psi}{m_0^2}$, i.e., in terms
of ${\cal L}_{NJL}$. Thus, $H$ is interpreted as the bound
$H \sim < \bar\psi \psi >$.

In a more general and natural setup one should expect multiple bound
states,
$< \bar t^\prime t^\prime >$, $< \bar b^\prime b^\prime >$ ...,
and, therefore, multiple scalar states below the compositeness scale
\cite{luty,hashimoto1,DEWSB-HLP,burdman1,hunghybrid,multiH}.
Moreover, new heavy fermions embedded within multi-Higgs models
are more favorable when confronted with the data
\cite{4G2HDM1,4G2HDM2,4G2HDM3}.

Indeed, a wide ``spectrum" of DEWSB/NJL models were suggested in the past
two decades: models of top-condensation and models of new heavy quarks or new heavy leptons condensation, producing either a single Higgs composite
or multi-Higgs composites. These models can be divided into two categories:
\begin{itemize}
\item Models where the condensing fermions have a mass
of order the EW-scale (e.g., top-condensation models), for
which the cutoff is
$\Lambda \sim {\cal O}(10^{17})$ GeV.
\item Models with new heavy fermions of mass ${\cal O}(500)$ GeV
(e.g., 4th generation models),
having a compositeness scale as low as $\Lambda \sim {\cal O}(1)$ TeV.
\end{itemize}
However, there is one major caveat in
all fermion-condensation models of the ``conventional"
NJL type: the typical mass of the composite $< \bar\psi \psi >$
tends to lie in the range $m_\psi < m_{< \bar\psi \psi >} < 2 m_\psi$,
thus being too heavy to account for the recently discovered $\sim 125$ GeV Higgs-like particle.

To bypass this difficulty, we have proposed in \cite{ourhybrid} an alternative solution
for the TeV-scale DEWSB scenario,
which leads to a light SM-like Higgs with a mass of ${\cal O}(m_W)$.
In particular, as mentioned earlier, 
we suggest a hybrid DEWSB setup with new heavy quarks
and a cutoff/threshold of $\Lambda \sim {\cal O}({\rm few})$ TeV,
by adding an unconstrained (i.e., fundamental) scalar field ($\Phi_\ell$) at the
compositeness scale, where the super-critical 4-Fermi operators form an
additional heavy composite ($\Phi_h$).

\section{The low-energy hybrid 2HDM (h4G2HDM)}

As a toy framework, we consider a new chiral 4th generation doublet, assumed
to be charged under some new strong interaction that dynamically 
break EW symmetry.\footnote{The DEWSB mechanism proposed 
here can be generalized to the case of
non-sequential TeV-scale vector-like quarks.} 
The theory at the compositeness scale,
$\Lambda$, can then be parameterized
by adding to the light SM degrees of freedom the following
set of strongly coupled 4-Fermi terms:
\begin{eqnarray}
{\cal L}={\cal L}_{SM}(\Lambda) + G_{t^\prime} \bar Q_L^\prime t^\prime_R {\bar t}^\prime_R Q_L^\prime +
G_{b^\prime} \bar Q_L^\prime b^\prime_R {\bar b}^\prime_R Q_L^\prime
+ G_{t^\prime b^\prime} \left( \bar Q_L^\prime b^\prime_R {\bar t}^{\prime c}_R i \tau_2 Q_L^{\prime c} + h.c. \right) ~,
\label{4fermi}
\end{eqnarray}
where $Q_L^\prime = (t^\prime_L ~ b^\prime_L)^T$ and ${\cal L}_{SM}(\Lambda)$
stands for the bare SM
Lagrangian with a single
fundamental Higgs field, $\Phi_\ell$,
which is essentially responsible for
the origin of mass of the lighter fermions.
As described in the previous section,
the above Lagrangian/theory can be reproduced by introducing an auxiliary
Higgs doublet $\Phi_h$, which couples ONLY to the 4th generation quarks 
as follows:
\begin{eqnarray}
{\cal L}_{q^\prime}(\Lambda) =
g_{b^\prime}^0 \left( \bar Q_L^\prime \Phi_h b^\prime_R + h.c. \right) +
g_{t^\prime}^0 \left( \bar Q_L^\prime \tilde\Phi_h t^\prime_R + h.c. \right)
- (\mu_h^0)^2 \Phi_h^\dagger \Phi_h ~,
\label{aux1}
\end{eqnarray}
so that the full theory at $\Lambda$ is then described by:\footnote{We have added
a $\Phi_h - \Phi_\ell$ mixing term
$\propto (\mu_{h \ell}^0)^2$, which may arise e.g., from QCD-like instanton effects
associated with the underlying strong dynamics (see e.g., \cite{Hill1994,PQ1})
or from sub-critical couplings of the
fundamental Higgs to the 4th generation quarks. This term
explicitly breaks the $U(1)$ Peccei-Quinn (PQ) symmetry \cite{PQ},
which is otherwise possessed by the model, thus
avoiding the presence of a massless pseudoscalar in the spectrum.
Note that, in any realistic scenario we expect
$\mu_{h \ell}(\mu \sim m_W) \sim {\cal O}(m_W)$ and,
since this is the only term which breaks the
PQ symmetry, it evolves only logarithmically under the RGE
so that, at the compositeness scale, we remain with
$\mu_{h\ell}^0 \equiv \mu_{h \ell}(\mu \sim \Lambda) \sim {\cal O}(m_W)$.
Therefore, since
$\mu_{h/\ell}^0 \equiv \mu_{h/\ell}(\mu \sim \Lambda) \sim {\cal O}(\Lambda)$,
we expect $(\mu_{h \ell}^0)^2/(\mu_{h}^0)^2 \sim {\cal O}(m_W^2/\Lambda^2) \ll 1$.}
\begin{eqnarray}
{\cal L}(\Lambda) = {\cal L}_{SM}(\Lambda) +
{\cal L}_{q^\prime}(\Lambda)
+ (\mu_{h \ell}^0)^2 \left( \Phi_h^\dagger \Phi_\ell + h.c. \right) ~.
\label{aux2}
\end{eqnarray}

Thus, integrating out $\Phi_h$, we recover the 4-Fermi Lagrangian with:
\begin{eqnarray}
G_{t^\prime} = \frac{(g_{t^\prime}^0)^2}{(\mu_h^0)^2}  ~,~
G_{b^\prime} = \frac{(g_{b^\prime}^0)^2}{(\mu_h^0)^2}  ~,~
G_{t^\prime b^\prime} = -\frac{g_{t^\prime}^0 g_{b^\prime}^0}{(\mu_h^0)^2} ~,
\label{Gterms}
\end{eqnarray}
plus additional subleading interaction
terms between the light Higgs and the new heavy quarks
with Yukawa couplings of
${\cal O} \left( \frac{(\mu_{h \ell}^0)^2}{(\mu_{h}^0)^2} \cdot g_{t^\prime/b^\prime}^0 \right)$ (see
\cite{ourhybrid}). In particular, $\Phi_h$ is now viewed as a composite
of the form
$\Phi_h \sim g_{t^\prime} < \bar Q_L^\prime t^\prime_R > + g_{b^\prime} < \bar Q_L^\prime b^\prime_R >$.

We thus end up with a specific 2HDM, where one (composite) doublet, $\Phi_h$, couples
to the new 4th generation fermions (which acquire their mass dynamically)
and one fundamental-like doublet (at $\Lambda$), $\Phi_\ell$,
which may be viewed as a pseudo-Goldstone of the underlying theory (see below) and
which is responsible for the mass generation of the SM fermions and for the CKM flavor structure.

At energies below $\Lambda$, $\Phi_h$ acquires a kinetic term as well as large self interactions
from the heavy fermion loops (as we will see below, the $\Phi_\ell$ quartic term does not
receive such large corrections), and the theory behaves
like the 4G2HDM of \cite{4G2HDM1}, with the scalar potential:
\begin{eqnarray}
V_{h4G2HDM}(\Phi_h,\Phi_\ell)  &=& \mu_\ell^2 \Phi_\ell^\dagger \Phi_\ell + \mu_h^2 \Phi_h^\dagger \Phi_h
-  \mu_{h \ell}^2 \left( \Phi_h^\dagger \Phi_\ell + h.c. \right)
+ \frac{1}{2} \lambda_\ell \left( \Phi_\ell^\dagger \Phi_\ell \right)^2 \nonumber \\
&+&
\frac{1}{2} \lambda_h \left( \Phi_h^\dagger \Phi_h \right)^2
\lambda_3 \left( \Phi_h^\dagger \Phi_h \right) \left( \Phi_\ell^\dagger \Phi_\ell \right)
+ \lambda_4 \left( \Phi_h^\dagger \Phi_\ell \right) \left( \Phi_\ell^\dagger \Phi_h \right)~,
\label{V}
\end{eqnarray}
where all the above mass terms and quartic couplings run as a function
of the energy scale $\mu$, as dictated
by the RGE for this model.\footnote{The most
general 2HDM potential also includes the
quartic couplings $\lambda_{5,6,7}$, which, in our
model, are absent at any scale.}
Note that the stability condition for the above potential
reads $\lambda_\ell,\lambda_h > 0$ and
$\sqrt{\lambda_\ell \lambda_h} > -\lambda_3 - \lambda_4$.

The 4-Fermi theory is now matched at $\Lambda$ to our h4G2HDM, by
solving the RGE of the model with the compositeness boundary conditions:
\begin{eqnarray}
g_{q^\prime}(\Lambda) \to \infty ~~,~~ \lambda_{h,3,4}(\Lambda) \to \infty ~~,~~
\lambda_h(\Lambda)/g_{q^\prime}^4(\Lambda) \to 0 ~~,~~
\lambda_{3,4}(\Lambda)/g_{q^\prime}^2(\Lambda) \to 0 ~, \label{cc}
\end{eqnarray}
where $q^\prime = t^\prime,b^\prime$. That is,
the composite theory is effectively a strongly coupled                                                            Higgs-Yukawa and Higgs-quartic systems, while
$\lambda_\ell (\mu \to \Lambda) \to \lambda_\ell^{(0)}$,
where $\lambda_\ell^{(0)}$ is a free parameter of the model.
The Higgs mass parameters (obtained after minimizing the above potential,
see \cite{ourhybrid}) are given by:
\begin{eqnarray}
\mu_{\ell}^2 \simeq \frac{\mu_{h \ell}^2}{t_\beta} - \frac{v^2}{2} c_\beta^2 \lambda_\ell ~~,~~
\mu_{h}^2 \simeq t_\beta \mu_{h \ell}^2 - \frac{v^2}{2} s_\beta^2 \lambda_h  ~,
\label{muterms}
\end{eqnarray}
where $t_\beta=v_h/v_\ell$, $s_\beta,c_\beta=\sin\beta,\cos\beta$
and it is understood that the quartic couplings are evaluated at $\mu \sim v$,
i.e., $\lambda_{h} = \lambda_{h}(\mu \sim v)$ and
$\lambda_\ell=\lambda_{\ell}(\mu \sim v)$, and we also have
$\mu_{h \ell}(m_W) \sim m_W$.

\section{The h4G2HDM and the 125 GeV Higgs}

The dominant RGE in our model are given approximately by (taking for simplicity
$g_{t^\prime}=g_{b^\prime} \equiv g_{q^\prime}$):
\begin{eqnarray}
{\cal D}g_{q^\prime} \approx 6 g_{q^\prime}^3  ~~,~~
{\cal D}\lambda_{h} \approx  4 \lambda_{h} \left( 3 \lambda_{h}+ 6 g_{q^\prime}^2 \right) - 24 g_{q^\prime}^4 ~, \label{RGE3}
\end{eqnarray}
where ${\cal D} \equiv 16 \pi^2 \mu \frac{d}{d\mu}$.
With the compositeness boundary conditions in Eq.~\ref{cc}, the above RGE's have a simple analytic solution:
\begin{eqnarray}
g_{q^\prime} \left(\mu\right) = \sqrt{\frac{4\pi^2}{3ln \frac{\Lambda}{\mu}}} ~~~,~~~
\lambda_{h} \left(\mu\right) = \frac{4\pi^2}{3ln \frac{\Lambda}{\mu}} \label{solutions} ~.
\end{eqnarray}

Thus, using $m_{q^\prime} = \frac{v_h \cdot g_{q^\prime}(\mu =m_{q^\prime})}{\sqrt{2}}$, we
can obtain the compositeness scale $\Lambda$ as a function of $m_{q^\prime}$ and $t_\beta$:
\begin{eqnarray}
\Lambda \approx m_{q^\prime} \cdot exp \left( \frac{2\pi^2\left( s_\beta v \right)^2}{3 m_{q^\prime}^2} \right)~,
\end{eqnarray}
so that for $m_{q^\prime} \sim {\cal O}(500)$ GeV and
$\tan\beta \sim {\cal O}(1)$ we obtain
$\Lambda \sim 1-1.5$ TeV.
This is many orders of magnitudes smaller than the cutoff in the top-condensation
scenario: $\Lambda \sim m_t \cdot exp\left(\frac{16\pi^2v^2}{9m_{t}^2} \right) \sim
10^{17}$ GeV (obtained by solving the SM-like RGE
for $g_t$: ${\cal D}g_{t} \approx \frac{9}{2} g_{t}^3$),
which, therefore, introduces a severe fine-tuning problem.

The physical scalar masses are given by:
$m_A^2 = m_{H^+}^2 = \frac{\mu_{h \ell}^2}{s_\beta c_\beta}$ and
\begin{eqnarray}
m_{h,H}^2 = \left( m_1^2 + m_2^2 \mp \sqrt{(m_1^2 - m_2^2)^2 + 4 \mu_{h \ell}^4} \right)/2 ~,
\end{eqnarray}
where (see also Eq.~\ref{muterms}):
\begin{eqnarray}
m_1^2 &\simeq&  \mu_{h}^2 + 3 s_\beta^2 v^2 \lambda_h /2 \simeq t_\beta \mu_{h \ell}^2 + s_\beta^2 v^2 \lambda_h ~, \\
m_2^2 &\simeq&  \mu_{\ell}^2 + 3 c_\beta^2 v^2 /2 \lambda_\ell \simeq \mu_{h \ell}^2 / t_\beta +
c_\beta^2 v^2 \lambda_\ell ~. \label{m1m2}
\end{eqnarray}

The Higgs mixing angle, which is defined by: $h = \cos\alpha \cdot {\rm Re}(\Phi_\ell^0) -
\sin\alpha \cdot {\rm Re}(\Phi_h^0)$ and
$H = \cos\alpha \cdot {\rm Re}(\Phi_h^0) +
\sin\alpha \cdot {\rm Re}(\Phi_\ell^0)$, is given by:
\begin{eqnarray}
\tan 2\alpha \simeq \left(
\cot 2 \beta -
\frac{v^2 \left( s_\beta^2 \lambda_h - c_\beta^2 \lambda_\ell \right)}{2 \mu_{h \ell}^2} \right)^{-1} ~.
\end{eqnarray}

\begin{figure}[htb]
\begin{center}
\includegraphics[scale=0.5]{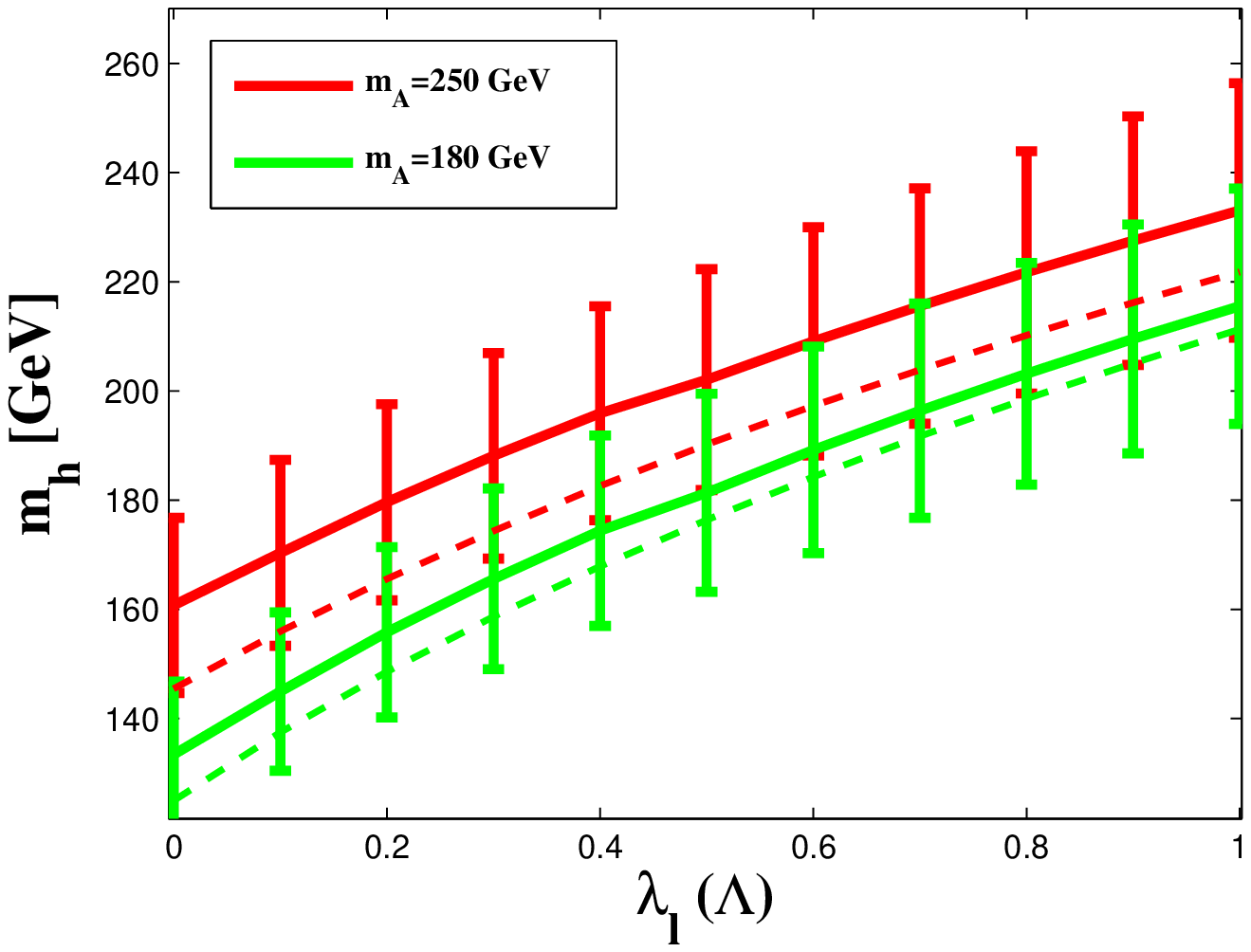}
\includegraphics[scale=0.5]{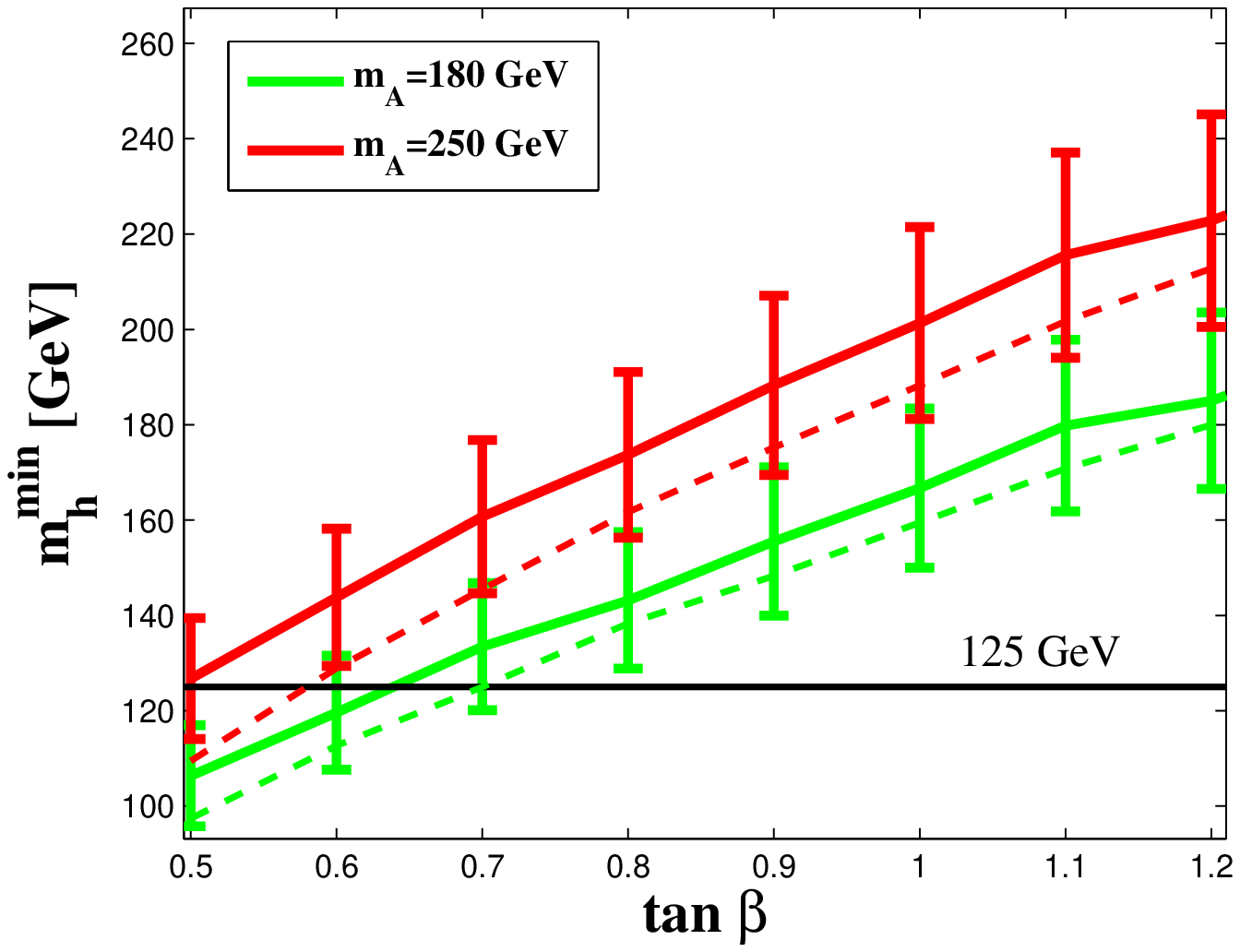}
\end{center}
\caption{\emph{Left: $m_h$ as a function of $\lambda_\ell(\Lambda)$, for
$\tan\beta=0.7$. Right: The minimal value of $m_h$, obtained
by choosing $\lambda_\ell(\Lambda)=0$ (see text),
as a function
of $\tan\beta$. Both plots are  for
$m_{q^\prime} = 400$ GeV and for $m_{A} = 180$ and $250$ GeV.
The approximate analytic solutions are shown by solid lines and
exact results (obtained from a full RGE analysis)
without errors by the dashed lines.}
\label{fig2}}
\end{figure}

Solving the RGE and evaluating the Higgs masses, we find that a light Higgs requires $\lambda_\ell(\mu=\Lambda) \to 0$, as demonstrated
in Fig.~\ref{fig2}, in which case the fundamental-like doublet
has a vanishing quartic term at $\Lambda$ and is, therefore, not
the SM doublet, but should rather be viewed as a pseudo-Goldstone boson
of the underlying strong dynamics.
In particular, for $\lambda_\ell(\Lambda) \to 0$
and $\tan\beta \sim {\cal O}(1)$, we obtain
$m_h \sim m_A/\sqrt{2}$. Thus,
for $m_A \sim 200-300$ GeV, we get $m_h \sim 125$ GeV
($\pm 10\%$), see Fig.~\ref{fig2}.
Furthermore, the above solution
corresponds to a small Higgs mixing angle of $\alpha \sim {\cal O}(10^0)$,
so that the light 125 GeV Higgs, $h$, is mostly the fundamental state.

The mass of the heavy CP-even Higgs, which is mostly the composite state,
is given by $m_H \sim v \sqrt{\lambda_h/2}$,
which for $m_{q^\prime} \sim {\cal O}(500)$ GeV is: $m_H \sim 500 \pm 100$ GeV.

Finally, the range of values for the free parameters $\tan\beta$ and $m_A$, which
gives a viable light Higgs candidate in our model, i.e., $\tan\beta \sim {\cal O}(1)$
and $m_A \sim 200-300$ GeV, also reproduce all the measured 125 GeV Higgs signals, 
as shown in \cite{ourhybrid}.

To summarize, we have introduced a hybrid mechanism for DEWSB,
where the compositeness scale is of order of a few TeV.
The model has new heavy quarks which acquire dynamical masses
and which form a heavy composite scalar. A fundamental-like scalar
is added at the compositeness scale and is responsible for the SM's flavor
structure and for the mass generation of the lighter fermions.
The EW symmetry is broken by combining
the Higgs mechanism with the NJL mechanism.
This allows us to get a viable 125 GeV Higgs candidate, which is
mostly fundamental and, therefore, protected from large $q^\prime$ loops.
Our model is consistent with all currently measured 125 GeV Higgs signals as well
as with EW precision data. The other low-energy Higgs states are
a charged scalar $H^+$ and a pseudo-scalar $m_A$ - both with a mass
in the range $m_A, m_{H^+} \sim 200 -300$ GeV, and a heavy
CP-even Higgs with a mass around 500 GeV.

\end{document}